\documentclass[a4paper,12pt]{article}
\usepackage{amsmath}
\usepackage{amssymb}
\usepackage{latexsym}
\newcommand{\be}{\begin{eqnarray}}
\newcommand{\ee}{\end{eqnarray}}
\begin{document}
\begin{titlepage}

\begin{centering}
\vspace{.3in}
{\Large{\bf Decoupling of the general scalar field mode and the solution space for Bianchi type I and V cosmologies coupled to perfect fluid sources}}
\\

\vspace{.5in} {\bf  T. Christodoulakis$^*$\footnote{tchris@cc.uoa.gr},
\quad Th. Grammenos$^\dag$\footnote{thgramme@uth.gr},
\quad Ch. Helias$^*$,
\\ P.G. Kevrekidis$^\ddag$\footnote{kevrekid@math.umass.edu},
\quad A. Spanou$^\S$}\\

\vspace{0.3in}
{\it $^*$ University of Athens, Physics Department\\
157 84 Athens, Greece\\
$^\dag$  University of Thessaly, Dept. of Mechanical \& Industrial Engineering\\
383 34 Volos,
Greece\\
$^\ddag$  University of Massachusetts, Department of Mathematics and Statistics,\\ Amherst MA 01003-4515, USA\\
$^\S$ National Technical University of Athens, Physics Department,\\
    157 73    Athens, Greece\\
}
\end{centering}

\vspace{0.7in}
\begin{abstract}
\par\noindent
The scalar field degree of freedom in Einstein's plus Matter field equations is decoupled for Bianchi type I
and V general cosmological models. The source, apart from the minimally coupled scalar field with arbitrary potential $V(\Phi)$,
is provided by a perfect fluid obeying a general equation of state $p = p(\rho)$. The resulting ODE is, by an appropriate
choice of final time gauge affiliated to the scalar field, reduced to 1st order, and then the system is completely integrated
for arbitrary choices of the potential and the equation of state.
\end{abstract}

\end{titlepage}
\newpage
\numberwithin{equation}{section}
\section{Introduction}
In the past two decades there has been a growing interest in scalar field cosmological models primarily due to the
prominent importance of scalar fields for inflationary scenarios \cite{Guth}. The importance of the coupling between a scalar
field and the gravitational field has been further stressed by Madsen \cite{Madsen}, who has shown that it can have non-trivial
consequences for the spontaneous breaking of gauge symmetries. A dynamical systems approach has been extensively used
in the study of scalar field cosmologies and their asymptotic behavior \cite{Belinskii}. See also  \cite{Wainwright}  for a concise review.

As far as exact solutions of scalar field cosmologies are
concerned, Burd and Barrow \cite{Burd} have studied homogeneous
but anisotropic Bianchi models of types III and VI (as well as
Kantowski-Sachs models) and have found exact solutions. Lidsey,
and Aguirregabiria et al \cite{Lidsey} have found exact solutions
for Bianchi type I models. Feinstein and Ibanez \cite{Feinstein}
have found exact solutions for Bianchi models of type III and VI,
Moss and Wright, Madsen, and Abreut, Crawford and Mimoso
\cite{Moss} have studied exact solutions in the setting of
conformal scalar field cosmologies.  Paul \cite{Paul} has obtained
exact solutions of a higher derivative theory in the presence of
an interacting scalar field. The discovery of the BTZ black hole
has motivated the study of analytic solutions in the context of
scalar field cosmology in (2+1) dimensions \cite{analytic}, while
Russo \cite{Russo} has obtained the general solution for a scalar
field cosmology in d dimensions with exponential potentials and a
flat Robertson-Walker metric. An early work on exponential
potentials has been done by Salopek and Bond \cite{salopek} while
recent treatments are given by Kehagias and Kofinas
\cite{kofinas2005} and  Neupane \cite{neupane}.

Cosmological models containing both a fluid and a scalar field have also been studied. Chimento and Jakubi \cite{Chimento}
have given exact solutions of scalar field cosmologies with a perfect fluid and a viscous fluid respectively. Mendez \cite{Mendez}
has obtained an exact solution for the case of an imperfect fluid in a FRW spacetime. In the so called scaling scalar field cosmologies,
the energy density due to the scalar field is proportional to the energy density of the perfect fluid. Thus a number of spatially flat,
isotropic models in which the energy density of the scalar field is proportional to that of the perfect fluid have been investigated \cite{isotropic}.
Billyard, Coley and van den Hoogen \cite{Billyard1} have studied the stability of these scaling solutions within the class of spatially
homogeneous cosmological models with a barotropic fluid matter content. Furthermore, they have studied the qualitative behavior
of spatially homogeneous models with a barotropic fluid and a non-interacting scalar field with an exponential potential in the class of
Bianchi type B models \cite{Billyard2}. Saha \cite{Saha} has obtained exact solutions for a Bianchi type I model with a perfect fluid
and dark energy content, while Chimento and Cossarini have studied exact solutions in 1+1 dimensions using an isotropic
perfect fluid source \cite{Cossarini}.

In a significant paper, Hawkins and Lidsey \cite{Hawkins}, have
shown that, for a flat FRW geometry, the dynamics of scalar field
cosmologies with a perfect fluid matter content can be described
by the non-linear, Ermakov-Pinney equation (which, in turn, leads
to tantalizing analogies with the dynamics of Bose-Einstein
condensates), while an early work of Barrow also deserves mention
in this context. Exact solutions have been obtained in this
description \cite{exact}. Another kind of decoupling of the scalar
field degree of freedom has been initiated in \cite{decoupling},
where a Robertson-Walker background geometry minimally coupled to
the scalar field has been investigated. By use of integrals of the
motion and of the Klein-Gordon equation in the quadratic
constraint equation, a single, higher-order non-linear
differential equation for the scalar field was obtained.

In the present work we generalize this decoupling for the case of
general Bianchi type I and V geometries in the presence of a
general perfect fluid source. The advantage is that the resulting
ODE is fully integrated and this is achieved for arbitrary choices
of the scalar field potential and the fluid equation of state. The
paper is organized as follows. In section 2 the geometry and the
matter description as well as the governing equations for the
system are presented. In section 3 the decoupling of the scalar
field degree of freedom is performed for Bianchi type I, the
reduction of the resulting ODE is given and the system is
completely integrated. The corresponding calculations for Bianchi
type V are given in section 4. Finally, the conclusions and a
discussion of the results obtained are presented in section 5.

Throughout we use geometrized units, i.e. $c = 8\pi G = 1$, while $g_{\mu\nu}$ has the signature $(-, +, +, +)$.

\section{The governing equations and the matter content}

Our starting point is the line element for a spatially homogeneous geometry:
\begin{equation}\label{2.1}
ds^2=[N^\alpha (t)N_\alpha (t)-N^2(t)]\,dt^2+
2N_\alpha(t)\sigma_{i}^{\alpha}(x)\,dx^i dt+
\gamma_{\alpha\beta}(t)\sigma_{i}^{\alpha}(x)\sigma_{j}^{\beta}(x)\,dx^idx^j
\end{equation}
where $\sigma_i^\alpha (x)$ are the basis one-forms satisfying
$\sigma_{i,j}^{\alpha}(x)-\sigma_{j,i}^{\alpha}(x)
=2C_{\beta\gamma}^{\alpha}\sigma_{j}^{\beta}(x)\sigma_{i}^{\gamma}(x)$,
(with $C_{\beta\gamma}^{\alpha}$ the structure constants of the
Lie algebra of the corresponding three-\break dimensional isometry
group of motion acting simply transitively on the spatial
hypersurfaces of simultaneity), $N_\alpha(t)$ is the shift vector,
$N(t)$ is the lapse function, and $\gamma_{\alpha\beta}(t)$ is the
scale factor matrix of the Bianchi type examined.

As explained in \cite{Korfiatis}, there are special general
coordinate transformations mixing space and time in the new space
coordinates, whose effect on the line element (\ref{2.1}) is
described by
\begin{equation}\label{2.2}
\begin{split}
& \tilde{\gamma}_{\alpha\beta}(t)=\Lambda_{\alpha}^{\mu}\Lambda_{\beta}^{\nu}\gamma_{\mu\nu}(t),\\
& \tilde{N}(t)=N(t)\\
& \tilde{N}_{\alpha}(t)=\Lambda_{\alpha}^{\beta}(t)\,[N_{\beta}(t)+P^{\rho}(t)\gamma_{\rho\beta}(t)]
\end{split}
\end{equation}
where $\Lambda_{\nu}^{\mu}(t)$, $P^{\rho}(t)$  satisfy
\begin{equation}\label{2.3a}
\Lambda_{\mu}^{\alpha}C_{\beta\gamma}^{\mu}=\Lambda_{\beta}^{\rho}\Lambda_{\gamma}^{\sigma}C_{\rho\sigma}^{\alpha}
\end{equation}
\begin{equation}\label{2.3b}
2P^{\mu}C_{\mu\nu}^{\alpha}\Lambda_{\beta}^{\nu}=\dot{\Lambda}_{\beta}^{\alpha}
\end{equation}
the overdot denoting differentiation with respect to time. Due to
(\ref{2.3a}), $\Lambda^{\alpha}_{\beta}(t)$ belongs to the
automorphism group of the Bianchi type in question and
transformations (\ref{2.2}) describe the gauge freedom of the
emanating system of Einstein's field equations in the case of
vacuum. So, the three arbitrary functions of time contained in
$\Lambda_{\beta}^{\alpha}(t)$ and $P^\alpha(t)$ can be used to
simplify the line element (\ref{2.1}) and, thus, also the
aforementioned equations. Automorphisms induced by general
coordinate transformations have also been considered in
\cite{Samuel}, while the rigid gauge symmetries have been analyzed
in \cite{Coussaert}. Time-dependent automorphisms, seen as tangent
space transformations, have also been considered previously
\cite{Jantz}, \cite{Uggla},while the first ( known to us)
reference of the relevance of Automorphisms to a systematic
analysis of Bianchi Cosmologies goes back to 1962 \cite{Heck}.

The above result also holds true when the matter content is such
that the linear constraints $G_{i}^{0}$ do not acquire an extra
term  $T_{i}^{0}$. For orthogonal perfect fluids and scalar fields
depending only on time, $T_{i}^{0}$ is indeed zero and therefore
we can use the gauge freedom (\ref{2.2}-\ref{2.3b}) to diagonalize
$\gamma_{\alpha\beta}(t)$ without loosing generality. Therefore,
the scale factor matrix is taken to be
\begin{equation}\label{2.4}
\gamma_{\alpha\beta}(t)=\mbox{diag}(a^2, b^2, c^2)
\end{equation}
while for the shift vector we have  $N_{\alpha}(t)=0$. Our choice of time is specified by the gauge
condition $N=\sqrt{\mbox{det}\gamma_{\mu\nu}}$, which is frequently used due to the simplification of the Einstein
tensor and whose importance for the decoupling is essential.

The matter content is a minimally coupled scalar field with an arbitrary potential $V(\Phi)$, thus having an energy-momentum tensor
\begin{equation}\label{2.5}
T_{\mu\nu}^{(1)}=\Phi_{,\mu}\Phi_{,\nu}-\frac{1}{2}g_{\mu\nu}(g^{\kappa\lambda}\Phi_{,\kappa}\Phi_{,\lambda}+2V(\Phi))
\end{equation}
and a perfect fluid part
\begin{equation}\label{2.6}
T_{\mu\nu}^{(2)}=(p+\rho)u_{\mu}u_{\nu}+pg_{\mu\nu}
\end{equation}
where $u^{\mu}$ is the unit four-velocity vector and a general
equation of state $p=p(\rho)$ is adopted. The governing Einstein's
Field Equations are taken to be
\begin{equation}\label{2.7}
R_{\nu}^{\mu}-\frac{1}{2}\delta_{\nu}^{\mu}R=T_{\nu}^{\mu}
\end{equation}
where  $T_{\mu\nu}=T_{\mu\nu}^{(1)}+T_{\mu\nu}^{(2)}$, while the scalar field equation (Klein-Gordon) is
\begin{equation}\label{2.8}
\frac{1}{\sqrt{-g}}\partial_{\mu}(\sqrt{-g}g^{\mu\nu}\partial_{\nu}\Phi)-V^{\prime}(\Phi)=0
\end{equation}
the prime, from now onwards, denoting differentiation with respect to the argument.

The ``equation of motion'' for the perfect fluid is the conservation of its energy-momentum tensor:
\begin{equation}\label{2.9}
T_{\nu;\mu}^{(2)\mu}=0
\end{equation}
(The scalar field energy-momentum tensor is separately conserved by virtue of the Klein-Gordon equation).

\section{Decoupling of the scalar degree and the solution space for Bianchi Type I}

The basis one-forms are
$\sigma_{i}^{\alpha}(x)=\delta_{i}^{\alpha}$ and, thus, with
(\ref{2.4}), zero shift and the chosen time gauge, the initial
form of the metric is given by
\begin{equation}\label{3.1}
g_{\mu\nu}=
\begin{pmatrix}
  -a^2b^2c^2 & 0 & 0 & 0 \\
  0 & a^2 & 0 & 0 \\
  0 & 0 & b^2 & 0 \\
  0 & 0 & 0 & c^2
\end{pmatrix}
\end{equation}
Spatial homogeneity implies that $\Phi=\Phi(t)$  and  $\rho=\rho(t)$. The nonzero components of the Einstein
tensor  $G_{\nu}^{\mu}$, and the energy-momentum tensors $T_{\nu}^{(1)\mu}$, $T_{\nu}^{(2)\mu}$  (all multiplied
by $a^2b^2c^2$) are given by
\begin{eqnarray}
G_0^0 &=& -\frac{\dot{a}\dot{b}}{ab}-\frac{\dot{a}\dot{c}}{ac}-\frac{\dot{b}\dot{c}}{bc}\label{3.2a}\\
\nonumber\\
G_1^1 &=& \frac{\dot{a}\dot{b}}{ab}+\frac{\dot{b}^2}{b^2}+\frac{\dot{a}\dot{c}}{ac}+\frac{\dot{b}\dot{c}}{bc}
+\frac{\dot{c}^2}{c^2}-\frac{\ddot{b}}{b}-\frac{\ddot{c}}{c}\label{3.2b}\\
\nonumber\\
G_2^2 &=& \frac{\dot{a}^2}{a^2}+\frac{\dot{a}\dot{b}}{ab}+\frac{\dot{a}\dot{c}}{ac}+\frac{\dot{b}\dot{c}}{bc}
+\frac{\dot{c}^2}{c^2}-\frac{\ddot{a}}{a}-\frac{\ddot{c}}{c}\label{3.2c}\\
\nonumber\\
G_3^3 &=& \frac{\dot{a}^2}{a^2}+\frac{\dot{a}\dot{b}}{ab}+\frac{\dot{b}^2}{b^2}+\frac{\dot{a}\dot{c}}{ac}
+\frac{\dot{b}\dot{c}}{bc}-\frac{\ddot{a}}{a}-\frac{\ddot{b}}{b}\label{3.2d}
\end{eqnarray}
\begin{eqnarray}
T_0^{(1)0} &=& -a^2b^2c^2\, V(\Phi)-\frac{1}{2}\dot{\Phi}^2\label{3.3a}\\
\nonumber\\
T_1^{(1)1} &=& T_2^{(1)2}=T_3^{(1)3}=-a^2b^2c^2\, V(\Phi)+\frac{1}{2}\dot{\Phi}^2\label{3.3b}\\
\nonumber\\
T_0^{(2)0} &=& -a^2b^2c^2\rho(t)\label{3.3c}\\
\nonumber\\
T_1^{(2)1} &=& T_2^{(2)2}= T_3^{(2)3}=a^2b^2c^2\,p(\rho(t))\label{3.3d}
\end{eqnarray}

Due to the equalities (\ref{3.3b}) and (\ref{3.3d}), we can subtract the corresponding Einstein's equations
$G_{\nu}^{\mu}=T_\nu^{(1)\mu}+T_\nu^{(2)\mu}$, i.e. form the differences (all multiplied by $a^2b^2c^2$)\break
$G_2^2-G_1^1=T_2^{(1)2}+T_2^{(2)2}-T_1^{(1)1}-T_1^{(2)1}$  and
$G_3^3-G_1^1=T_3^{(1)3}+T_3^{(2)3}-T_1^{(1)1}-T_1^{(2)1}$ and get the following equations involving only the scale factors
\begin{equation}\label{3.4}
\frac{\dot{a}^2}{a^2}-\frac{\dot{b}^2}{b^2}-\frac{\ddot{a}}{a}+\frac{\ddot{b}}{b}=0
\end{equation}
\begin{equation}\label{3.5}
\frac{\dot{a}^2}{a^2}-\frac{\dot{c}^2}{c^2}-\frac{\ddot{a}}{a}+\frac{\ddot{c}}{c}=0
\end{equation}
These equations provide the following integrals of motion:
\begin{equation}\label{3.6a}
b(t)=e^{\lambda t}a(t)
\end{equation}
\begin{equation}\label{3.6b}
c(t)=e^{\mu t}a(t)
\end{equation}
leaving us with just one undetermined scale factor. At this stage the Klein-Gordon equation (\ref{2.8}) becomes
\begin{equation}\label{3.7}
-V^{\prime}(\Phi)-\frac{e^{-2(\lambda+\mu)t}\ddot{\Phi}(t)}{a^6(t)}=0
\end{equation}
Now, the conservation of the fluid energy-momentum tensor
(continuity equation)
gives
\begin{equation}\label{3.8}
3\frac{\dot{a}}{a}+\frac{\dot{\rho}}{p(\rho)+\rho}+\lambda+\mu=0
\end{equation}

It is straightforward to check that the time derivative of the
quadratic constraint equation $G_0^0=T^{(1)0}_{0}+T^{(2)0}_{0}$ is
identically satisfied by virtue of the remaining spatial equation
$G_1^1=T^{(1)1}_{1}+T^{(2)1}_{1}$, the Klein-Gordon equation and
the continuity equation (solved for $\ddot{a}(t)$,
$\ddot{\Phi}(t)$, $\dot{\rho}(t)$), as expected from the
consistency between the aforementioned constraint with the spatial
Einstein and the Matter equations. Therefore, the equations to be
solved are the Klein-Gordon equation (\ref{3.7}), the continuity
equation (\ref{3.8}) and the constraint equation which, upon
multiplication by $-2a^6 e^{2(\lambda+\mu)t}$ reads:
\begin{equation}\label{3.9}
6\left(\frac{\dot{a}}{a}\right)^2+4(\lambda+\mu)\frac{\dot{a}}{a}+2\lambda\mu-
\dot{\Phi}^2-2e^{2(\lambda+\mu)t}a^6[V(\Phi)+\rho]=0
\end{equation}
In order to have a closed form for the integral of the continuity eq. (\ref{3.8}), it is convenient to use the parametrization
\begin{equation}\label{3.10}
p(\rho)=\frac{g(\rho)}{g^\prime(\rho)}-\rho
\end{equation}
(the prime denoting differentiation with respect to the argument)
through the use of which one obtains the integral
\begin{equation}\label{3.11}
\rho=h[\rho_0 a^{-3}e^{-(\lambda+\mu)t}]
\end{equation}
with $h$ being the inverse function to $g$, i.e. satisfying $h(g(x))=x$.\\
>From the Klein-Gordon equation (\ref{3.7}) the scale factor and its logarithmic derivative is expressed in terms of the scalar
field $\Phi(t)$ and its derivatives as well as $V(\Phi)$:
\begin{equation}\label{3.12}
a=\left[-\frac{e^{-2(\lambda+\mu)t}\ddot{\Phi}}{V^\prime(\Phi)}\right]^{1/6}
\end{equation}
\begin{equation}\label{3.13}
\frac{\dot{a}}{a}=\frac{1}{6}
\left(
\frac{\dddot{\Phi}}{\ddot{\Phi}}-\frac{V^{\prime\prime}(\Phi)}{V^\prime(\Phi)}\dot{\Phi}-2(\lambda+\mu)
\right)
\end{equation}
The promised decoupling of the scalar field dynamics from the geometry occurs upon inserting
(\ref{3.11}-\ref{3.13}) into (\ref{3.9}). The result is the following non-linear, third order ODE for $\Phi(t)$:
\begin{equation}\label{3.14}
\begin{split}
\left(
\frac{\dddot{\Phi}}{\ddot{\Phi}}-\frac{V^{\prime\prime}(\Phi)}{V^\prime(\Phi)}\dot{\Phi}
\right)^2 -
&4(\lambda+\mu)^2 + 12\lambda\mu -6\dot{\Phi}^2\\
 + & 12\frac{\ddot{\Phi}}{V^\prime(\Phi)}V(\Phi)
 +12\frac{\ddot{\Phi}}{V^\prime(\Phi)}h\left[\rho_0\left(-\frac{\ddot{\Phi}}{V^\prime(\Phi)}\right)^{-1/2}\right]=0
\end{split}
\end{equation}
At this stage, any solution to this equation determines, through
(\ref{3.11}) and (\ref{3.12}) a corresponding solution to the full
Einstein plus Matter system whose entire space of solution is
therefore attained from the solution space of (\ref{3.14}). Of
course, the price paid, for the moment, is the non-linearity in
the highest time derivative, which has also been raised to third
order. Furthermore, normally one would expect that the arbitrary
functions h, V need first to be specified before hoping to
actually get a solution. Nevertheless, it is quite interesting
that further reduction of the order of (\ref{3.14}), and
subsequent complete integration of the whole system, is possible.
To this end, we first observe that many terms in (\ref{3.14}),
namely all the nontrivial except the 4th and 5th, are functions of
the combination
$-\displaystyle{\frac{\ddot{\Phi}}{V^\prime(\Phi)}}$. We  thus
define:
\begin{equation}\label{3.15}
\chi=-\frac{\ddot{\Phi}}{V^\prime(\Phi)}
\end{equation}
and write (\ref{3.14}) as
\begin{equation}\label{3.16}
\left(
\frac{\dot{\chi}}{\chi}
\right)^2-4(\lambda+\mu)^2+12\lambda\mu-6f(\chi)-12\chi h(\rho_0\chi^{-1/2})=0
\end{equation}
where
\begin{equation}\label{3.17}
f(\chi)\equiv\dot{\Phi}^2+2\chi V(\Phi)
\end{equation}
implicitly reflects the arbitrariness in choosing $V(\Phi)$. Now (\ref{3.16}) can be integrated and, by
judicious choices for $f$, $h$, even give $\chi(t)$ in closed form. Suppose $\chi(t)$ does indeed solve (\ref{3.16}).
Then, multiplying (\ref{3.15}) by $2\dot{\Phi}V^\prime(\Phi)$  and using  (\ref{3.17}) we get the first order
linear differential equation for $f(t)$:
\begin{equation}\label{3.18}
\frac{df(t)}{dt}=2\dot{\chi}(t)V(t)
\end{equation}
and consequently $\Phi(t)$ is given by
\begin{equation}\label{3.19}
\frac{d\Phi(t)}{dt}=\pm \sqrt{f(\chi(t))-2\chi(t)V(t)}
\end{equation}
Finally, a change of time variable from $t$ to $\chi$ (defined in
(\ref{3.15})) permits the presentation of the entire space of
solutions to the system under consideration in closed form.
Indeed, considering $V$, $f$, $\Phi$ as functions of $\chi$ one
gets from (\ref{3.18}-\ref{3.19}) respectively
\begin{equation}\label{3.20}
\frac{df}{d\chi}=2V(\chi)
\end{equation}
\begin{equation}\label{3.21}
\frac{d\Phi(\chi)}{d\chi}\dot{\chi}(t)=\pm \sqrt{f(\chi)-2\chi
V(\chi)}
\end{equation}
where the time derivative of $\chi$ is given by the constraint (\ref{3.16}).
The integration of these two equations is trivial, yielding $f$, $\Phi$ as:
\begin{equation}\label{3.22}
f(\chi)=\sigma+2\int V(\chi)\, d\chi
\end{equation}
\begin{equation}\label{3.23}
\Phi(\chi)=\kappa \pm \int\frac{1}{\chi}\sqrt{\frac{f(\chi)-2\chi
V(\chi)} {4(\lambda+\mu)^2-12\lambda\mu+6f(\chi)+12\chi h (\rho_0
\chi^{-1/2})}}\, d\chi
\end{equation}
The line element in the new time $\chi$ reads:
\begin{equation}\label{3.24}
\begin{split}
ds^2 &= -\frac{d\chi^2}{\chi[4(\lambda+\mu)^2-12\lambda\mu+6f(\chi)+12\chi h(\rho_0 \chi^{-1/2})]}\\
&\\
& +[e^{-2(\lambda+\mu)t(\chi)}\chi]^{1/3}\, dx^2
+[e^{(4\lambda-2\mu)t(\chi)}\chi]^{1/3}\, dy^2
+[e^{(-2\lambda+4\mu)t(\chi)}\chi]^{1/3}\, dz^2
\end{split}
\end{equation}
with $t(\chi)$ given by the integral form of (\ref{3.16})
\begin{equation}\label{3.25}
t(\chi)=\pm
\int\frac{1}{\chi\sqrt{4(\lambda+\mu)^2-12\lambda\mu+6f(\chi)+12\chi
h (\rho_0\chi^{-1/2})}}\,d\chi
\end{equation}
while the density and pressure are given as:
\begin{equation}\label{3.26}
\rho(\chi)=h(\rho_0\chi^{-1/2}) \qquad\qquad p(\chi)=\rho_0 \chi^{-1/2}h^\prime (\rho_0 \chi^{-1/2})-h(\rho_0 \chi^{-1/2})
\end{equation}
the prime denoting differentiation with respect to the argument.
Quite independently of the way these solutions were obtained, one
can straightforwardly check ( through,say, a symbolic computing
facility) that they do satisfy all ten Einstein's Equations, the
generalized K-G and the continuity equation. Furthermore, since no
extra ansatz has been involved in the process of integration of
the system (\ref{2.7}-\ref{2.9}), equations
(\ref{3.22}-\ref{3.26}) represent the full space of solutions to
the Einstein plus Matter system considered. The functions $V$, $h$
can be freely specified to obtain special case solutions.

\section{Decoupling of the scalar degree and the solution space for the Bianchi Type
V}

In this case the basis one-forms are  $\sigma^1=e^{-x}dy$,
$\sigma^2=e^{-x}dz$, $\sigma^3=dx$. The $G_1^0=0$ Einstein
equation (due to spatial homogeneity $\Phi=\Phi(t)$ and
$\rho=\rho(t)$ and therefore there is no corresponding component
of the matter tensor) implies $c=ab$. Thus the initial metric is
taken as:
\begin{equation}\label{3.27}
g_{\mu\nu}=\begin{pmatrix}
  -a^3b^3 & 0 & 0 & 0 \\
  0 & ab & 0 & 0 \\
  0 & 0 & a^2 e^{-2x} & 0 \\
  0 & 0 & 0 & b^2 e^{-2x}
\end{pmatrix}
\end{equation}
The nonzero components of the Einstein tensor  $G_{\nu}^{\mu}$, and the energy-momentum
tensors  $T_{\nu}^{(1)\mu}$, $T_{\nu}^{(2)\mu}$, all multiplied by $a^3 b^3$, are given by
\begin{eqnarray}
G_0^0 &=& 3a^2b^2-\frac{\dot{a}^2}{2a^2}-\frac{2\dot{a}\dot{b}}{ab}-\frac{\dot{b}^2}{2b^2}\label{3.28a}\\
\nonumber\\
G_1^1 &=& a^2b^2+\frac{3\dot{a}^2}{2a^2}+\frac{2\dot{a}\dot{b}}{ab}+\frac{3\dot{b}^2}{2b^2}-
\frac{\ddot{a}}{a}-\frac{\ddot{b}}{b}\label{3.28b}\\
\nonumber\\
G_2^2 &=& a^2b^2+\frac{\dot{a}^2}{a^2}+\frac{2\dot{a}\dot{b}}{ab}+\frac{2\dot{b}^2}{b^2}-
\frac{\ddot{a}}{2a}-\frac{3\ddot{b}}{2b}\label{3.28c}\\
\nonumber\\
G_3^3 &=& a^2b^2+\frac{2\dot{a}^2}{a^2}+\frac{2\dot{a}\dot{b}}{ab}+\frac{\dot{b}^2}{b^2}-
\frac{3\ddot{a}}{2a}-\frac{\ddot{b}}{2b}\label{3.28d}
\end{eqnarray}
\begin{eqnarray}
T_0^{(1)0} &=& -a^3b^3\, V(\Phi)-\frac{1}{2}\dot{\Phi}^2\label{3.29a}\\
\nonumber\\
T_1^{(1)1} &=& T_2^{(1)2}=T_3^{(1)3}=-a^3b^3\, V(\Phi)+\frac{1}{2}\dot{\Phi}^2\label{3.29b}\\
\nonumber\\
T_0^{(2)0} &=& -a^3b^3\rho(t)\label{3.29c}\\
\nonumber\\
T_1^{(2)1} &=& T_2^{(2)2}= T_3^{(2)3}=a^3b^3\,p(\rho(t))\label{3.29d}
\end{eqnarray}

The situation is similar to the Type I case, and thus forming the difference
(multiplied by $a^3b^3$) $G_2^2-G_1^1=T_2^{(1)2}+T_2^{(2)2}-T_1^{(1)1}-T_1^{(2)1}$
we get the following equation involving only the scale factors
\begin{equation}\label{3.30a}
-\frac{\dot{a}^2}{2a^2}+\frac{\dot{b}^2}{2b^2}+\frac{\ddot{a}}{2a}-\frac{\ddot{b}}{2b}=0
\end{equation}
which can be integrated yielding
\begin{equation}\label{3.30b}
b(t)=e^{\lambda t}a(t)
\end{equation}
The Klein-Gordon equation (\ref{2.8}) becomes
\begin{equation}\label{3.31}
-V^\prime (\Phi)-\frac{e^{-3\lambda t}\ddot{\Phi}(t)}{a^6 (t)}=0
\end{equation}
while the conservation of the fluid energy-momentum tensor
(continuity equation) gives
\begin{equation}\label{3.32}
3\frac{\dot{a}}{a}+\frac{\dot{\rho}}{p(\rho)+\rho}+\frac{3}{2}\lambda=0
\end{equation}
Again, the only other equation to be solved is the constraint equation which reads
\begin{equation}\label{3.33}
6\left(\frac{\dot{a}}{a}\right)^2+6\lambda\frac{\dot{a}}{a}+\lambda^2-6e^{2\lambda t}a^4-
\dot{\Phi}^2-2e^{3\lambda t}a^6(V(\Phi)+\rho)=0
\end{equation}
where the fourth term is the only non trivial  difference from the corresponding equation (\ref{3.9})
and its presence is due to the non-vanishing curvature of the spatial slice.

Integrating (\ref{3.32}) (in the parameterization (\ref{3.10})) and solving (\ref{3.31})
for $a(t)$, we obtain the following results for the matter density $\rho(t)$, the scale factor $a(t)$
and its logarithmic derivative
\begin{equation}\label{3.34}
\rho=h\left(\rho_0 a^{-3}e^{-\frac{3}{2}\lambda t}\right)
\end{equation}
\begin{equation}\label{3.35}
a=\left(-\frac{e^{-3\lambda t}\ddot{\Phi}}{V^{\prime}(\Phi)}\right)^{1/6}
\end{equation}
\begin{equation}\label{3.36}
\frac{\dot{a}}{a}=\frac{1}{6}\left(\frac{\dddot{\Phi}}{\ddot{\Phi}}
-\frac{V^{\prime\prime}(\Phi)}{V^{\prime}(\Phi)}\dot{\Phi}-3\lambda\right)
\end{equation}
Use of these equations in the quadratic constraint equation (\ref{3.33}) (multiplied by $-2e^{3\lambda t}a^6$) yields
\begin{equation}\label{3.37}
\begin{split}
\left(\frac{\dddot{\Phi}}{\ddot{\Phi}}
-\frac{V^{\prime\prime}(\Phi)}{V^{\prime}(\Phi)}\dot{\Phi}\right)^2-3\lambda^2
& -36\left(-\frac{\ddot{\Phi}}{V^\prime(\Phi)}\right)^{2/3} - 6\dot{\Phi}^2+12\frac{\ddot{\Phi}}{V^{\prime}(\Phi)}V(\Phi)\\
&\\
& +12\frac{\ddot{\Phi}}{V^{\prime}(\Phi)}h \left(\rho_0
\left(-\frac{\ddot{\Phi}}{V^\prime (\Phi)}\right)^{-1/2}\right)=0
\end{split}
\end{equation}
which, with the same definition of $\chi$, translates into
\begin{equation}\label{3.38}
\left(\frac{\dot{\chi}}{\chi}\right)^2-3\lambda^2-36\chi^{2/3}-6f(\chi)-12\chi
h (\rho_0 \chi^{-1/2})=0
\end{equation}
By arguments completely analogous to the previous Type I case, the
final form of the solution is, in this case:
\begin{equation}\label{3.39}
f(\chi)=\sigma+2\int V(\chi)\, d\chi
\end{equation}
\begin{equation}\label{3.40}
\Phi(\chi)=\kappa \pm \int\frac{1}{\chi}\sqrt{\frac{f(\chi)-2\chi
V(\chi)} {3\lambda^2+36\chi^{2/3}+6f(\chi)+12\chi h (\rho_0
\chi^{-1/2})}}\, d\chi
\end{equation}
\begin{equation}\label{3.41}
\begin{split}
ds^2 &= -\frac{d\chi^2}{\chi[3\lambda^2+36\chi^{2/3}+6f(\chi)+12\chi h(\rho_0 \chi^{-1/2})]}\\
&\\
& +\chi^{1/3}\, dx^2
+e^{-\lambda t(\chi)-2x}\chi^{1/3}\, dy^2
+e^{\lambda t(\chi)-2x}\chi^{1/3}\, dz^2
\end{split}
\end{equation}
with $t(\chi)$ given by the integral form of (\ref{3.38})
\begin{equation}\label{3.42}
t(\chi)=\pm
\int\frac{1}{\chi\sqrt{3\lambda^2+36\chi^{2/3}+6f(\chi)+12\chi h
(\rho_0\chi^{-1/2})}}\,d\chi
\end{equation}
while the density and pressure are given as:
\begin{equation}\label{3.43}
\rho(\chi)=h(\rho_0\chi^{-1/2}) \qquad\qquad p(\chi)=\rho_0 \chi^{-1/2}h^\prime (\rho_0 \chi^{-1/2})-h(\rho_0 \chi^{-1/2})
\end{equation}\\The remarks at the end of the previous section
apply also here

\section{Discussion-Conclusions}

We have discussed the dynamics of a scalar field with an arbitrary
potential, minimally coupled to a general (anisotropic) Bianchi
Type I and V geometry, in the presence of a perfect fluid source
obeying a general equation of state. In the case of vacuum, the
rich structure of Outer Automorphisms for these two symmetry
groups entails the existence of some integrals of motion: Indeed,
consider the generators of the rigid symmetries (\ref{2.3a}), i.e.
the vector fields in the space of dependant variables
$X_{I}=\lambda^\rho_{I\alpha}\,\gamma_{\rho\beta}\,
\frac{\partial}{\partial\gamma_{\alpha\beta}}$ where
$\lambda^\alpha_\beta$ satisfy
$\lambda^\alpha_{I\rho}\,C^\rho_{\beta\gamma}=\lambda^\rho_{I\beta}\,C^\alpha_{\rho\gamma}+
\lambda^\rho_{I\gamma}\,C^\alpha_{\beta\rho}$ .
If one performs the appropriate change of variables
$\gamma_{\alpha \beta}\rightarrow z_{\alpha\beta}$ which brings one
(or more, if possible) generator into each canonical form, say
$\frac{\partial}{\partial z_{11}}$, then Einstein's field equations
written in the new variables do not explicitly depend on $z_{11}$.
Thus the system becomes of first order in the variable \.{z}$_{11}$ and
therefore an integral of motion arises.
\\
The initial choice of the time gauge in which the lapse is equal to the determinant
of the scale factor matrix has a twofold advantage: Firstly, it
enables the corresponding integrals of motion (in the presence of
the matter content chosen) to be revealed. Secondly, it makes the
Klein-Gordon equation purely algebraic in the scale factor
variables. As a result, when the utilization of the integrals of
motion has reduced the number of these variables to one, this
equation gives this last scale factor as a function of the second
derivative of the scalar field, the derivative of the potential
with respect to the scalar field, and of time. Substitution of
this form of the scale factor into the only equation remaining to
be solved, i.e. the quadratic constraint, results in a single ODE
for the scalar field (without any explicit time dependence!). The
utilization of a final time gauge adapted to the scalar field
enables the reduction of this equation to first order and
subsequently leads to the complete integration of the entire
system of Einstein plus Matter Field Equations. The description of
the space of solutions contains, in an integral form, the
arbitrary functions of the final time $\chi$, $V(\chi)$ and
$h(\rho_0 \chi^{-1/2})$. The presence of this twofold
arbitrariness corresponds to the fact that we have not specified
either the form of the potential (as a function of the scalar
field) or the equation of state. It is evident that prescribing
$V(\chi)$ and $h(\rho_0 \chi^{-1/2})$ implicitly corresponds to a
choice of potential form and equation of state. For example, the
choice $h = A(\rho_0 \chi^{-1/2})^{1+\gamma}$ gives through
(\ref{3.43}) $p =\gamma\rho$, i.e. the barotropic equation of
state. For the scalar field, if we take the particular case
$\lambda=\mu=\sigma= h(\rho_0 \chi^{-1/2}) = 0$ (i.e., flat
Robertson-Walker with no fluid) the choice
$V(\chi)=\displaystyle{\frac{C}{\sqrt{\chi}}}$ corresponds to the
functional form $V=\pm Ce^{\sqrt{3}(\kappa-\Phi)}$. If, on the
other hand, someone insists in prescribing $V(\Phi)$ and
$p(\rho):=p(h)$ then the situation must be dealt with in the
following manner: As far as the density is concerned, the equation
giving the pressure becomes the holonomic, first order
differential equation $p(h)=wh^\prime (w)-h(w)$ with $w$ standing
for $\rho_0\chi^{-1/2}$ which can be straightforwardly integrated.
For example, a barotropic equation $p:=\gamma\rho$ gives $\rho=
A(\rho_0\chi^{-1/2})^{1+\gamma}$. As far as the matter field is
concerned, the situation is somewhat more complicated as the
results of choosing a particular form $V(\Phi)$ are influenced by
the choice of the density $h$. As an example consider again the
case $\lambda = \mu =\sigma = h(\rho_0\chi^{-1/2}) = 0$ and an
arbitrary potential form $V(\Phi)$. Relaxing for a moment
(\ref{3.23}), we get the following Klein-Gordon equation
\[-\frac{V^{\prime}(\Phi)}{V(\Phi)}+\frac{6\chi}{6\chi^2 \Phi^{\prime \, 2}(\chi)}
[2\chi\Phi^{\prime\prime}(\chi)-6\chi^2\Phi^{\prime \, 3}(\chi)+3\Phi^{\prime}(\chi)]=0\]
which may be difficult to solve depending on the choice of $V(\Phi)$.
This is the price paid for insisting on prescribing $V(\Phi)$ and not $V(\chi)$ in which case the solution
would be given by (\ref{3.22}-\ref{3.26}) or (\ref{3.39}-\ref{3.43}) correspondingly. The particular choice
$V = V_0 e^{-\lambda\Phi}$ corresponds to the case considered in \cite{Russo} and the equation above can be
dealt with by choosing a new time $\chi = e^{3\tau /2}$, in which the equation becomes the following first
order in the derivative $\omega \equiv \displaystyle{\frac{d\Phi}{d\tau}}$:
\[48\frac{d\omega}{d\tau} - 96\omega^3 + 24\lambda\omega^2 - 9\lambda = 0\]

Finally, there are two cases which, at first sight, need to be
separately examined. The first concerns the case of a constant
potential $V(\Phi)\equiv V_{0}$, i.e. a cosmological constant
term. Then, $V^{\prime}(\Phi)=0$ and the definition of the final
time $\chi$ in equation (\ref{3.15}) seems to be precarious.
However, the Klein-Gordon eq. (\ref{3.7}) then implies that also
$\ddot{\Phi}=0$ and, surprisingly enough, $\chi$ does exist.
Indeed, substituting $V(\Phi)\equiv V_{0}$ in the solutions
(\ref{3.22}-\ref{3.26}) or (\ref{3.39}-\ref{3.43}) one can see
that the solution is still valid. This holds true even for
$V_{0}=0$.
The second case arises when the ratio
$\displaystyle{\frac{\ddot{\Phi}}{V^{\prime}(\Phi)}=-\frac{1}{c^{6}}}$, for,
then, the change of time from $t$ to $\chi$ is not valid. In this
case, the remaining scale factor is given by
$a(t)=ce^{-\frac{(\lambda+\mu)t}{3}}$ in the Type I case or
$a(t)=ce^{-\frac{\lambda t}{2}}$ in the Type V case. The
continuity equation ((\ref{3.8}) or (\ref{3.31}) respectively)
implies that the pressure and density are constants, say, $p_{0}$
and $\rho_{0}$. The remaining equations dictate $\Phi(t)=At+B$ and
$V(\Phi)=V_{0}$ and there exists a relation between
$c,p_{0},\rho_{0},A$ and $B$ due to the quadratic constraint.\\

The general properties of the space of solutions found for both
type I and type V cases, e.g. isotropization, attractors and
self-similarity, and quintessence, will be examined in a forthcoming
paper, in which also particular cases of high interest are to be
explicitly elaborated.  Possible applications of the method
exhibited here can be the cases of  $D + 1$ spatially homogeneous
spacetimes, in which the Outer Automorphisms are rich enough to
provide sufficient integrals of motion. Then the method will be
applicable if the matter content is such that the integrals of
motion persist.


\end{document}